\documentclass[a4paper]{jpconf}
\usepackage{graphicx}
\usepackage{bm}
\usepackage{amsmath,amssymb}

\begin{document}
\title{Elliptic Flow and Shear Viscosity of the Shattered Color Glass Condensate}

\author{Marco Ruggieri}
\address{Department of Physics and Astronomy, University of Catania, Via S. Sofia 64, I-95125 Catania}
\ead{marco.ruggieri@lns.infn.it}
\author{Francesco Scardina, Salvatore Plumari, Vincenzo Greco}
\address{Department of Physics and Astronomy, University of Catania, Via S. Sofia 64, I-95125 Catania}
\address{INFN-Laboratori Nazionali del Sud, Via S. Sofia 62, I-95123 Catania, Italy}

\begin{abstract}
In this talk, we report on our results about the computation of
the elliptic flow of the quark-gluon-plasma produced in relativistic heavy ion
collisions, simulating the expansion of the fireball by solving
the relativistic Boltzmann equation for the parton distribution function tuned at a fixed shear viscosity to entropy density
ratio $\eta/s$. 
We emphasize the role of saturation in the initial
gluon spectrum modelling the shattering of the color glass condensate, 
causing the initial distribution to be out of equilibrium.
We find that the saturation reduces the efficiency
in building-up the elliptic flow, even if the thermalization process is quite fast $\tau_{therm} \approx 0.8 \,\rm fm/c$.
and the pressure isotropization even faster $\tau_{isotr} \approx 0.3 \,\rm fm/c$.
The impact of the initial non-equilibrium manifests for non-central collisions 
and can modify the estimate of the viscosity respect to the assumption of full thermalization in $p_T$-space.
\end{abstract}

\section{Introduction}
In the last decade it has been reached a general consensus that Ultra-relativistic heavy-ion collisions (uRHICs) 
at the Relativistic Heavy-Ion Collider (RHIC) and the Large Hadron Collider (LHC) create a hot and dense 
strongly interacting quark and gluon plasma (QGP)~~\cite{STAR_PHENIX, ALICE_2011, Science_Muller,Fries:2008hs}. 
A main discovery has been that the QGP has a very small shear viscosity to entropy density, $\eta/s$,
which is more than one order of magnitude smaller than the one of water~\cite{Csernai:2006zz,Lacey:2006bc}, 
and close to the lower bound of $1/4\pi$ conjectured for systems at infinite strong coupling \cite{Kovtun:2004de}. 
A key observable to reach such a conclusion is the elliptic flow~\cite{Ollitrault},
$v_2 = \left\langle (p_x^2-p_y^2)/(p_x^2+p_y^2) \right\rangle$. 
In fact, the expansion of the created matter
generates a large anisotropy of the emitted particles that can be primarily measured by $v_2$.  
The origin of $v_2$ is the initial spatial eccentricity, 
$\epsilon_x=\langle y^2-x^2\rangle/\langle x^2+y^2 \rangle$ 
of the overlap region in non-central collisions, which is responsible for different
pressure gradients in the transverse plane thus favoring flow preferably 
along the $x$ direction rather than  $y$ direction.
The observed large $v_2$ is considered a signal of a very small $\eta/s$
because if $\eta/s$ was large then viscosity would damp flow, hence reducing anisotropy in
momentum space.
Calculations agree in indicating an average $\eta/s$ of the QGP lying in the range
$4\pi\eta/s \approx 1-3$~\cite{Romatschke:2007mq,Heinz,Song:2011hk,Schenke:2010nt,Niemi:2011ix,
Ferini:2008he, Xu:2008av,Xu:2007jv,Bratkovskaya:2011wp,Plumari_BARI,Ruggieri:2013bda,Uphoff:2014cba}.

Along with the existence of a deconfined QGP matter and the understanding of its properties, 
the uRHIC program offers the opportunity to verify the picture in which
the two colliding nuclei are described as two sheets of
Color Glass Condensate (CGC)~\cite{McLerran:1993ni}.
The CGC would be generated by the very high density of the gluon distribution
function at low $x$ (parton longitudinal momentum fraction), which triggers a saturation
of the distribution for $p_T$ below a saturation scale, $Q_s$. 
The determination of the shear viscosity $\eta/s$ of the QGP and the search for the CGC are related:
in fact the uncertainty on the initial condition translates into an uncertainty on the
theoretical estimate of $\eta/s$~\cite{Luzum:2008cw, Alver:2010dn,Song:2011hk,Adare:2011tg}.

In this talk we report our results \cite{Ruggieri:2013bda,Ruggieri:2013ova} about the computation of
the elliptic flow of the quark-gluon-plasma produced in relativistic heavy ion
collisions, simulating the expansion of the fireball by solving
the relativistic Boltzmann equation for the parton distribution function tuned at a fixed shear viscosity to entropy density
ratio $\eta/s$~\cite{Ruggieri:2013ova,Ferini:2008he,Plumari:2010ah,Plumari_BARI,Plumari:2012xz}. 
The advantage of using kinetic theory is that starting from a one-body
phase space distribution function $f(x,p)$, and not from the energy-momentum tensor $T^{\mu\nu}$, 
it is straightforward to initialize simulations from a non-equilibrium distribution function, 
while hydrodynamics relies on the gradient expansion of the
stress tensor and therefore is applicable only if initial deviations from
local equilibrium are small.

In this study we will consider several kinds of initial conditions,
two of them related to the shattering of the CGC which takes place after the collision.
For describing the latter we make use of the KLN 
model~\cite{Kharzeev:2004if,Hirano:2005xf,Drescher:2006pi,Drescher:2006ca,Hirano:2009ah},
which has been largely employed to study
the dynamics of HIC and the viscosity of the QGP within
hydrodynamical simulations~\cite{Song:2011hk,Luzum:2008cw,Alver:2010dn,Hirano:2009ah,
Shen:2013vja,Shen:2013cca,Song:2013qma}. 
The uncertainty in the initial condition 
translates into an uncertainty on $\eta/s$ of at a least a factor of two as estimated
by mean of several viscous hydrodynamical approaches \cite{Luzum:2008cw,Alver:2010dn,Song:2011hk,Adare:2011tg}. 
We point out that the implementation of the shattered CGC initialization
in hydrodynamics takes into account only the different space distribution
respect to a geometric Glauber model,
discarding the key and more peculiar feature of the damping
of the distribution for $p_T$ below the $Q_s$ saturation scale.
We have found by mean of kinetic theory
that this has a significant impact on the build-up of $v_2$,
hence on the determination of $\eta/s$.

In our study we neglect the initial time evolution of the
chromo-electric and chromo-magnetic fields (the glasma) produced immediately after the
collision. 
According to the commonly accepted picture of heavy ion collisions, 
our approach should be justified as soon as the initial strong fields decay 
into particle quanta.
Characteristic time for the glasma decay 
is of the order of $\tau_0\approx 1/Q_s$,
which turns out to be of the order of a fraction of fm/c~\cite{Ryblewski:2013eja,Gelis:2013rba,Fukushima:2013dma}. 
From this time on the field contribution to the pressure
is less important, and the dynamics is dominated by the scatterings among partons,
which justifies the use of kinetic theory to describe the sequent evolution of the system.
The advantage
to use kinetic theory from $\tau_0$ rather than hydro is that the nonequilibrium
initial distribution is not problematic as kinetic theory is built 
to study the evolution of a generic $f(x,p)$ distribution function. 

\section{Initial conditions for the simulations}
In this Section we specify the initial conditions we implement in our simulations.
The first one is based on the Glauber model,
see~\cite{Miller:2007ri} for a review, with an $\bm x$-space distribution given 
by the a standard mixture $0.85N_{part}+0.15N_{coll}$
and a $\bm p$-space thermalized spectrum in the transverse plane at a time $\tau_0 = 0.6\rm \, fm/c$
for the RHIC runs and $\tau_0 = 0.3$ fm/c for the LHC runs. 
Maximum temperature is assumed $T_0 = 0.34$ GeV for the Au-Au collisions at RHIC energy 
and $T_0 = 0.51$ GeV for the Pb-Pb collisions at LHC energy.
We also assume $y=\eta$ at $\tau = \tau_0$, where $\eta$ corresponds to 
space-time rapidity; we also assume
boost invariance in the longitudinal direction at $\tau=\tau_0$ which implies independence of the initial distribution on $y$.
Following the nomenclature introduced in~\cite{Ruggieri:2013bda,Ruggieri:2013ova} we will refer to this case as Th-Glauber.

The other two kinds of initial conditions we use in our simulations~\cite{Ruggieri:2013bda,Ruggieri:2013ova}
are based on a model of gluon production for the shattered CGC named 
factorized-KLN (fKLN in the following)~\cite{Kharzeev:2004if,Drescher:2006ca,Hirano:2009ah}, 
in which the initial spectrum is assumed to be 
\begin{equation}
\frac{dN}{dy d^2\bm x_T d^2 \bm p_T} =\kappa
\frac{p_A p_B }{p_T^2}
\int^{p_T} d^2\bm{k}_T
\alpha_S(Q^2)
\phi_A(x_1,k_T^2;\bm{x}_\perp)
\phi_B(x_2,(\bm{p}_T -\bm{k}_T)^2;\bm{x}_\perp)~;
\label{eq:densityPS}
\end{equation}
the coordinate space distribution
is obtained by integration of the above equation over $\bm p_T$. 
Here $(\bm p_T,y)$ correspond to transverse momentum and momentum rapidity of the
produced gluons respectively, and $x_{1,2} = p_T \exp(\pm y)/\sqrt{s}$.
In Eq.~\eqref{eq:densityPS} $p_{A,B}$ denote the probability to find one nucleon
at a given transverse coordinate, $p_A({\bm x}_\perp) = 1-
\left[1-{\sigma_{in}} T_A(\bm x_\perp)/A\right]^A$
where $\sigma_{in}$ is the inelastic cross section and  $T_A$ corresponds
to the usual thickness function of the Glauber model. 
The overall constant $\kappa$ in Eq.~\eqref{eq:densityPS} is fixed in order to
reproduce the experimental multiplicity in the most central collisions.

The specific form of the unintegrated gluon distribution function for gluons
$\phi_{A,B}$ is not important
here; it is enough to remind that it embeds saturation in the sense that 
$\phi_{A,B}$ as a function of $p_T$ is constant for $p_T< Q_s$,
where $Q_s$ corresponds to the saturation scale. 
Here we show results obtained within a parameter set
such that $\langle Q_s\rangle_{y=0} \approx 1$ GeV for the most central collisions
at longitudinal momentum fraction $x=0.01$.

Having described the fKLN model we are now ready to define the other two kinds of
initializations we use in simulations. Firstly we assume that both the initial 
$p_T$-spectrum and the coordinate space distribution are given by
Eq.~\eqref{eq:densityPS} and its integral over $\bm p_T$, which corresponds to the proper implementation of the fKLN model.
We call this initialization as fKLN initialization. 
This kind of initial condition is not implemented in hydro simulations because the initial
strong deviation from equilibrium cannot be studied within viscous hydrodynamics. 
In the fKLN initialization we assume $\tau_0=0.2$ fm/c because no assumption on equilibration
needs to be introduced; this time can be interpreted as the time needed for the strong
initial fields to decay to a parton liquid, as explained in the Introduction. 
On the other hand, we also consider another
initial condition which is close to the one usually implemented in hydro simulations,
in which we assume the coordinate space distribution given by the KLN model
and a $p_T$-thermalized spectrum. In this way we assume some thermalization occured in 
a short time, leaving at the same time the distribution in coordinate space unaffected.
We call this initialization as Th-fKLN initialization. For the latter we assume 
the same initialization times used for the Th-Glauber model. In the two cases we assume
$y=\eta$ at $\tau= \tau_0$.

\section{Kinetic theory at fixed $\eta/s$}
For the dynamical evolution of the initial conditions 
we employ transport theory as a base of a simulation code of the fireball expansion created
in relativistic heavy-ion collision~\cite{Ferini:2008he,Plumari:2010ah,Plumari_BARI,Plumari:2012xz}; 
we assume the system can be described by a gluon distribution function $f({\bm x}, {\bm p}, t)$
which evolves according to the Relativistic Boltzmann Transport (RBT) equation:
\begin{equation}
p_\mu\partial^\mu f = \int d\Gamma_2 d\Gamma_{1^\prime} d\Gamma_{2^\prime}(f_{1^\prime}f_{2^\prime} - f f_2)
|{\cal M}|^2\delta^4(p + p_2 - p_{1^\prime} - p_{2^\prime})~, 
\label{eq:BE}
\end{equation}
with $d^3 {\bm p}_{k} = 2 E_{k} (2\pi)^3 d\Gamma_k$, and ${\cal M}$ corresponds to the transition amplitude.
At variance with the standard use of transport theory we have developed
an approach that, instead of focusing on specific microscopic calculations of the scattering matrix, 
fixes the total cross section cell by cell in the grids in order to have a wished value of $\eta/s$. 
By means of this procedure we translate transport theory to hydrodynamical language since 
we use the Boltzmann equation to simulate
the dynamical evolution of a fluid which dissipates with a specified
shear viscosity, in analogy to what is done within hydrodynamical simulations.
Hence this approach offers a bridge between transport theory and hydrodynamics.

The advantage of the kinetic theory approach at fixed $\eta/s$, compared
to hydro simulations, is twofold: the initial condition does not need
to be close to thermal equilibrium, because we do not perform
a gradient expansion of the energy-momentum tensor. Moreover we do not
need to specify an ansatz for the deviations from equilibrium.
This approach to kinetic theory has been also considered in~\cite{Huovinen:2008te}
where it has been shown that transport theory at fixed $\eta/s$ reproduces
the results of viscous hydrodynamics for one-body observable like $T^{\mu\nu}$ or entropy density 
also in the limit in which the system is not in the dilute regime.
This is not surprising because looking at the Boltzmann collision integral in terms of viscosity
allows the analytical derivation of second order viscous hydrodynamics 
\cite{Grad,Denicol:2012es,Tsumura:2013uma,Bazow:2013ifa}.   
Despite the fact we consider only the $2\leftrightarrow 2$ processes 
to compute the collision integral, 
once the system is close to the
hydro regime with many collisions happening in a short time range, 
the detail of the single scattering is lost and what matters is only
the viscosity of the fluid and not the specific microscopic process producing
such viscosity. 
We have checked we work in a regime where the specific process is 
not relevant by changing the microscopic two-body scattering matrix from anisotropic to isotropic
and adjusting the total cross section to keep the same $\eta/s$. 
We have found that leaving unchanged all the other parameters,
the elliptic flow is not affected by this change for $p_T \leq 2.5$ GeV~\cite{greco-inpc}. 

Once $\eta/s$ is fixed, we compute the total cross section in each cell of the coordinate
space of our grid. The Chapmann-Enskog approximation 
supplies the relation between $\eta/s$ an cross section with quite good approximation, in agreement with the results 
obtained using the Green Kubo formula~\cite{Plumari:2012ep,Plumari:2012xz}:
\begin{equation}
\sigma_{tot}=\frac{1}{15}\frac{\langle p\rangle}{\rho \, g(a)} 
\frac{1}{\eta/s}=\frac{1}{15} \langle p\rangle \, \tau_{\eta}
~, 
\label{eq:sigma}
\end{equation}
which is valid for a generic differential cross section $d\sigma/dt \sim 1/(t-m_D^2)^2$~\cite{Plumari:2012ep}.
In the above equation $a=T/m_D$, with $m_D$ the screening mass regulating the angular dependence
of the cross section, while  $g(a)$ corresponds to a function whose analytic form can be found 
in \cite{Ruggieri:2013bda,Ruggieri:2013ova}. In the regime were viscous hydrodynamic applies 
(not too large $\eta/s$ and $p_T$)
the specific microscopic detail of the cross section is irrelevant and our approach 
is an effective way to employ transport theory to simulate a fluid at a given $\eta/s$.

The $\eta/s$ we use in our study \cite{Ruggieri:2013ova} is as follows:
in the plasma phase it is a constant, whose numerical value is fixed case by case;
we then implement a kinetic freezout by assuming that $\eta/s$ increases smoothly in a temperature range from the
plasma phase to a hadron phase values which is fixed by referring to the estimates 
in~\cite{Prakash:1993bt,Chen:2007xe,Demir:2008tr}. 
In this way we take into account scatterings in the hadron phase as well, which however give a very tiny
contribution to the collective flow because of the damping due to the larger viscosity.

\section{Thermalization}
\begin{figure}[t!]
\begin{center}
\includegraphics[width=6cm]{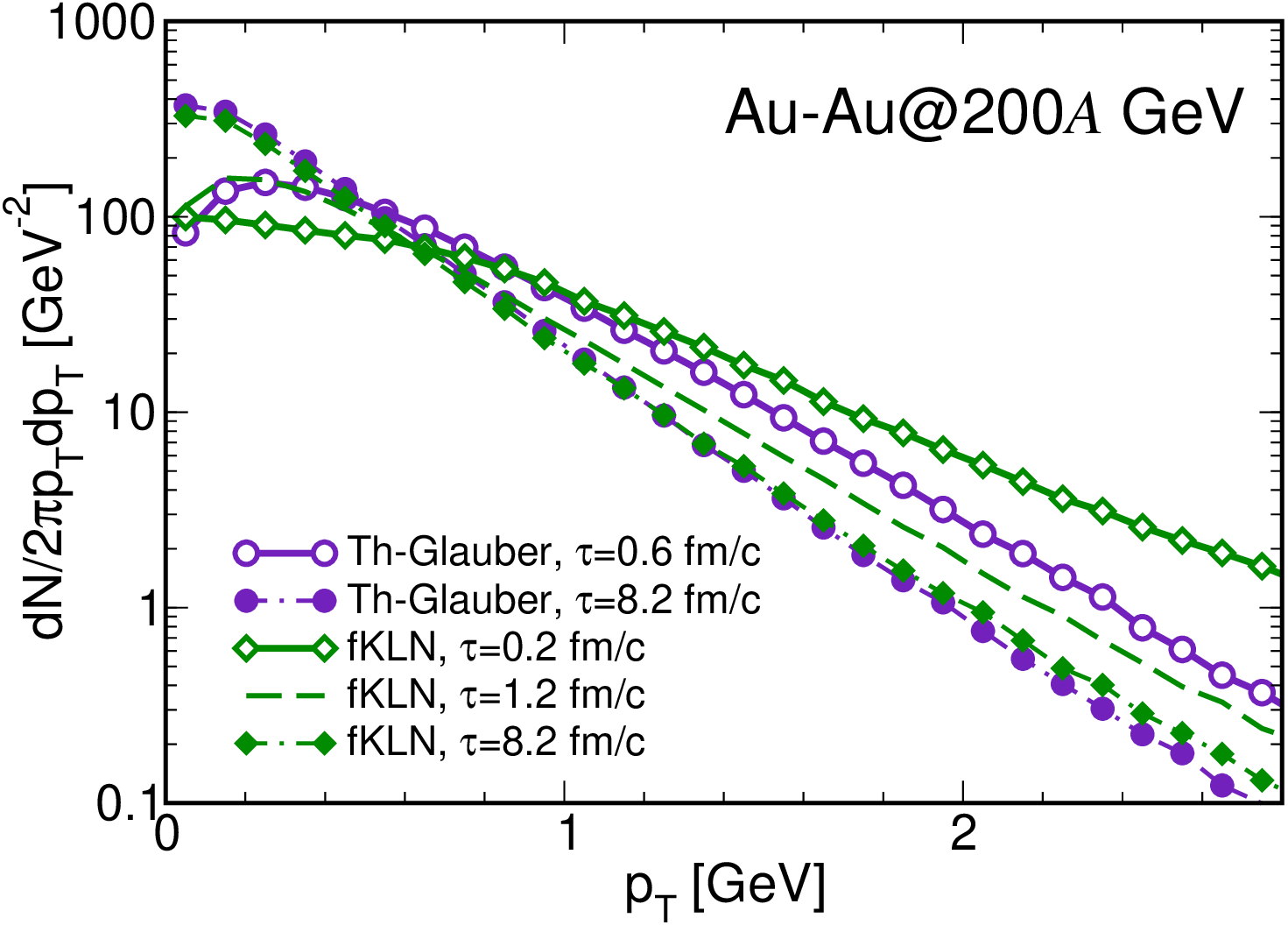}~~~\includegraphics[width=6cm]{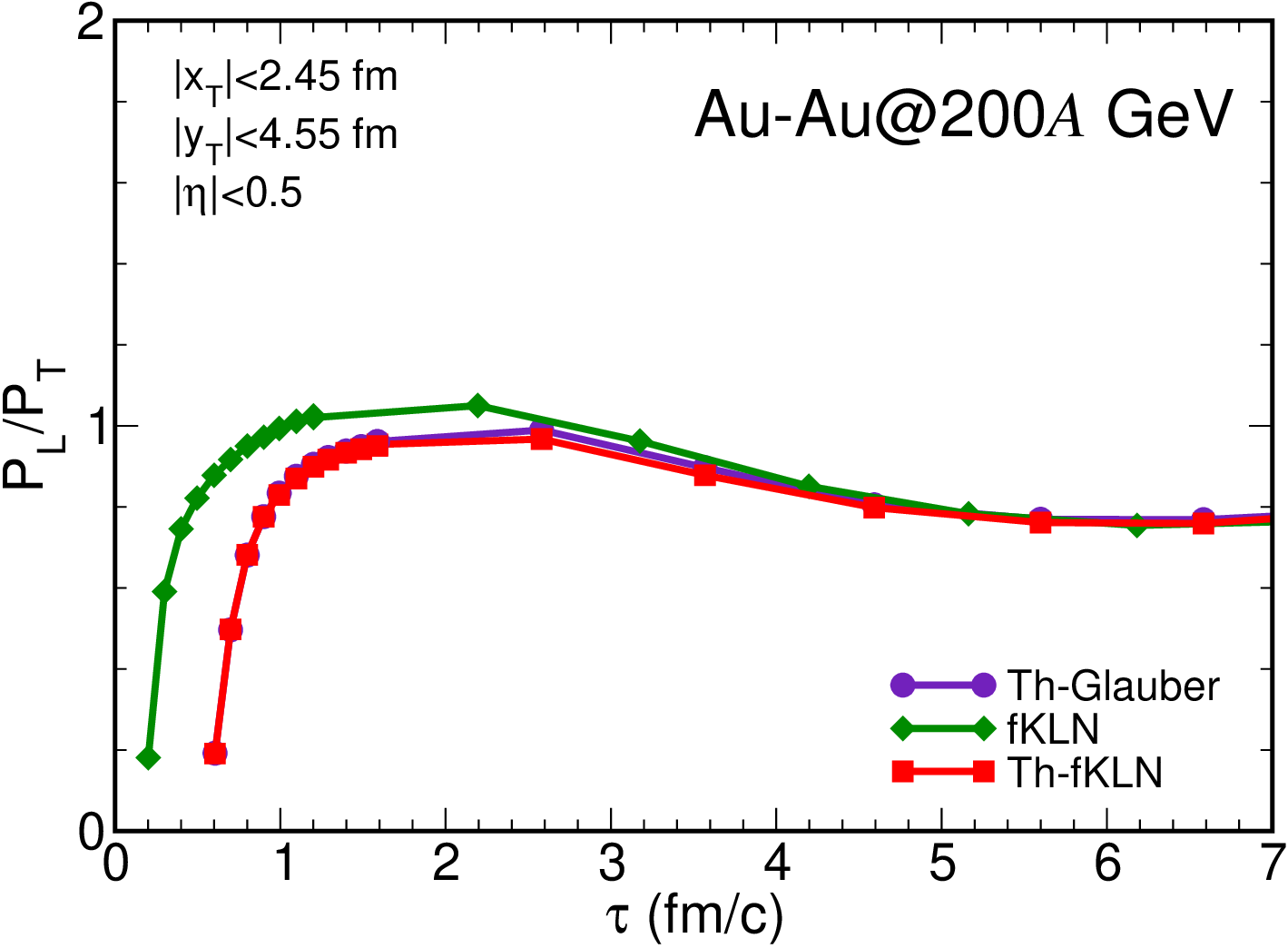}\\
\includegraphics[width=6cm]{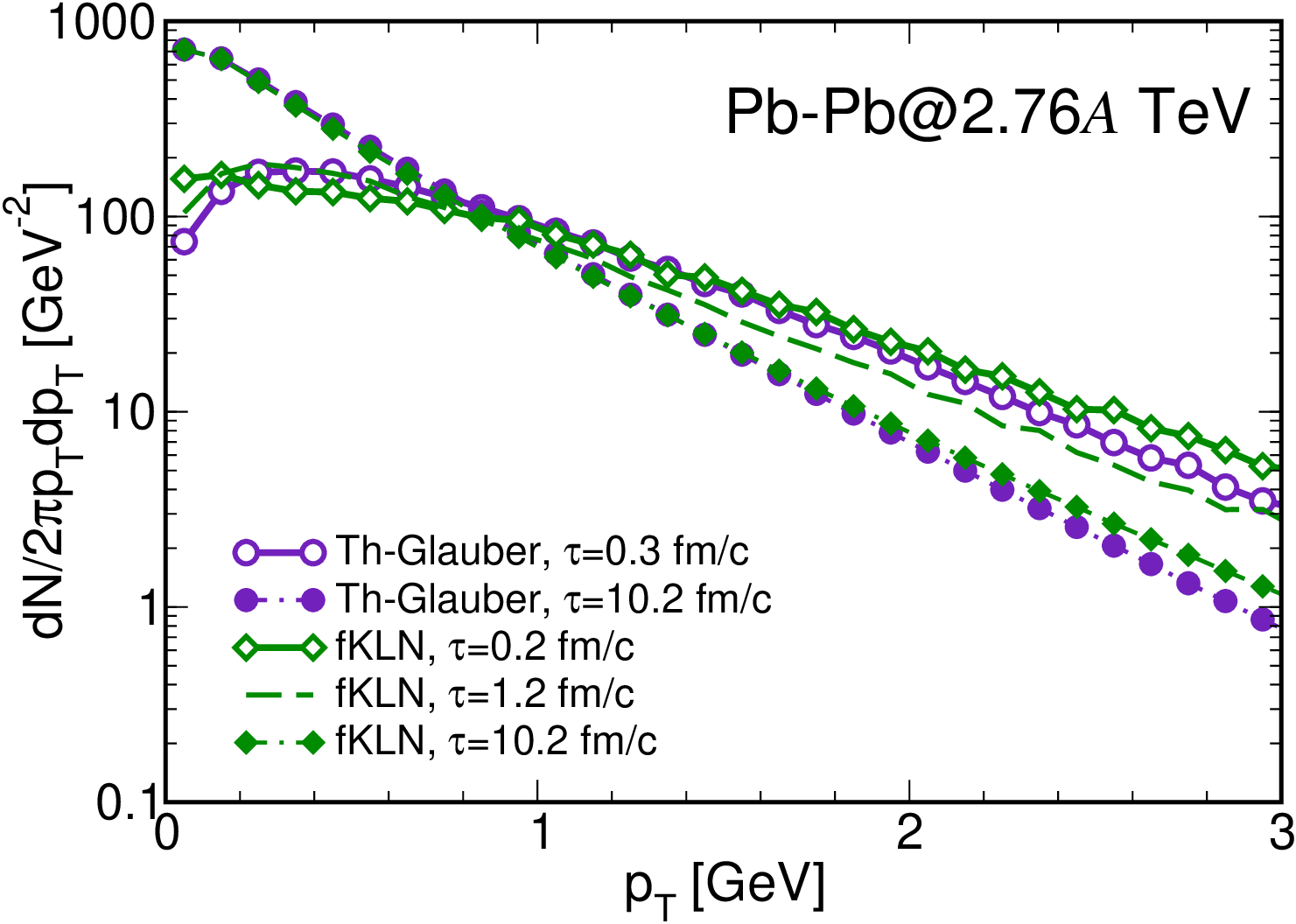}~~~\includegraphics[width=6cm]{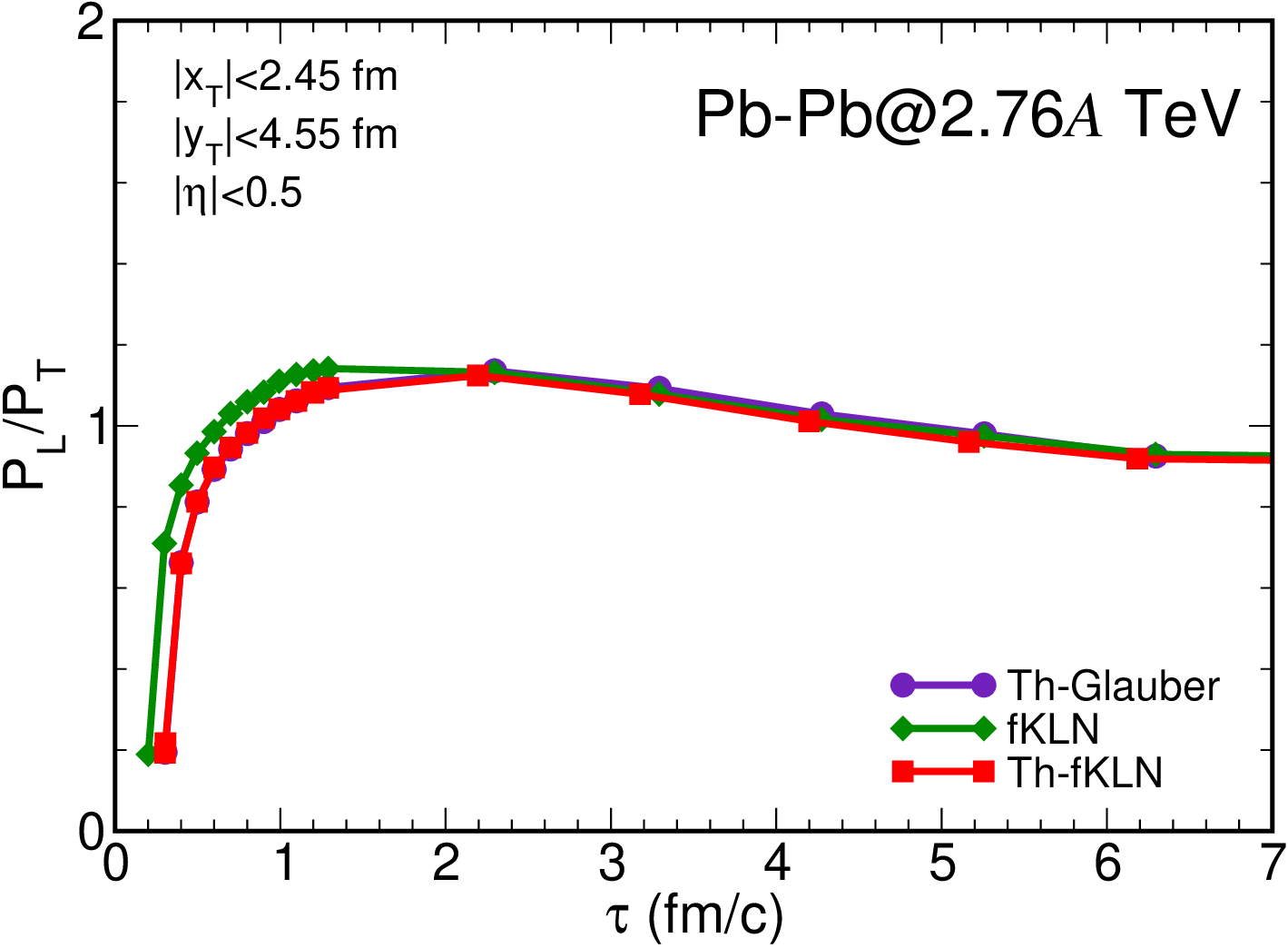}
\caption{\label{Fig:spettri_CGC2}
{\em Left panels:} Time evolution of spectra, for collisions at RHIC (upper left) and LHC (lower left) energies.
{\em Right panels:} Time evolution of $P_L/P_T$.  
In all panels $b=7.5$ fm and $4\pi\eta/s=1$.
The spectra for the Th-fKLN case are not shown because
we do not find visible deviations from the Th-Glauber 
initialization. Adapted from Ref.~\cite{Ruggieri:2013ova}.}
\end{center}
\end{figure}

In the upper left panel of Fig.~\ref{Fig:spettri_CGC2} we collect the initial spectra, $dN/2\pi p_T d p_T$,  
integrated over the momentum rapidity window $|y|<0.5$,
for the case of a Au-Au collision $\sqrt{s}=200 A$ GeV  
for the fKLN and Th-Glauber initial conditions at their respective initial
times $\tau_0$ and at the final time $\tau=8.2$ fm/c.
For the case of the Th-fKLN, we find that the spectra are the same as in the case
of the Th-Glauber, therefore we do not plot them in the figure. 
We have also shown by the dashed line the spectrum at $\tau=1.2$ fm/c for the case
of the fKLN initial condition. 
The results in Fig.~\ref{Fig:spettri_CGC2} are obtained with $4\pi\eta/s=1$.
We notice that initially the fKLN spectrum is quite far from a thermalized spectrum;
nevertheless the spectrum thermalizes in the transverse plane within $1$ fm/c,
since its $p_T$ dependence becomes exponential with a slope very similar to the Th-Glauber.
In~\cite{Ruggieri:2013ova} we have also checked thermalization is achieved in the
full momentum space within $\tau_{therm}\approx 0.8$ fm/c by computing the time evolution
of the quantity $T^* \equiv T\tau^{1/3}$, with $T=E/3N$
representing the temperature in the case of a thermalized system.
In the case of 1D dissipation-less expansion $T^*$ saturates to a constant as a function of time,
which we find to happen for the fKLN case within $\tau_{therm}$. 
Results for the LHC runs are shown on the lower left panel of Fig.~\ref{Fig:spettri_CGC2}.

On the upper right panel of Fig.~\ref{Fig:spettri_CGC2} we plot the ratio $P_L/P_T$ where $P_L$, $P_T$
correspond to the longitudinal and transverse pressure respectively. In the initial stage
of the collision there is a strong anisotropy due to the fact that $P_L\ll P_T$; however
the system efficiently removes the anisotropy in the case of small $\eta/s$, and in fact we find
that the time required to get isotropic pressure $P_L \approx P_T$ is $\tau_{iso}\approx 0.3$ fm/c.
In~\cite{Ruggieri:2013ova} we show that in case $\eta/s$ is quite large and comparable with the
perturbative QCD estimates, the system is not efficient in removing the initial anisotropy, which is
quite natural because the large viscosity damps the flow which instead is necessary to transfer 
momentum and equilibrate the pressures. 
Results for the LHC runs are shown on the lower right panel of Fig.~\ref{Fig:spettri_CGC2}.

\section{Elliptic flow}

\begin{figure}[t!]
\begin{center}
\includegraphics[width=7.5cm]{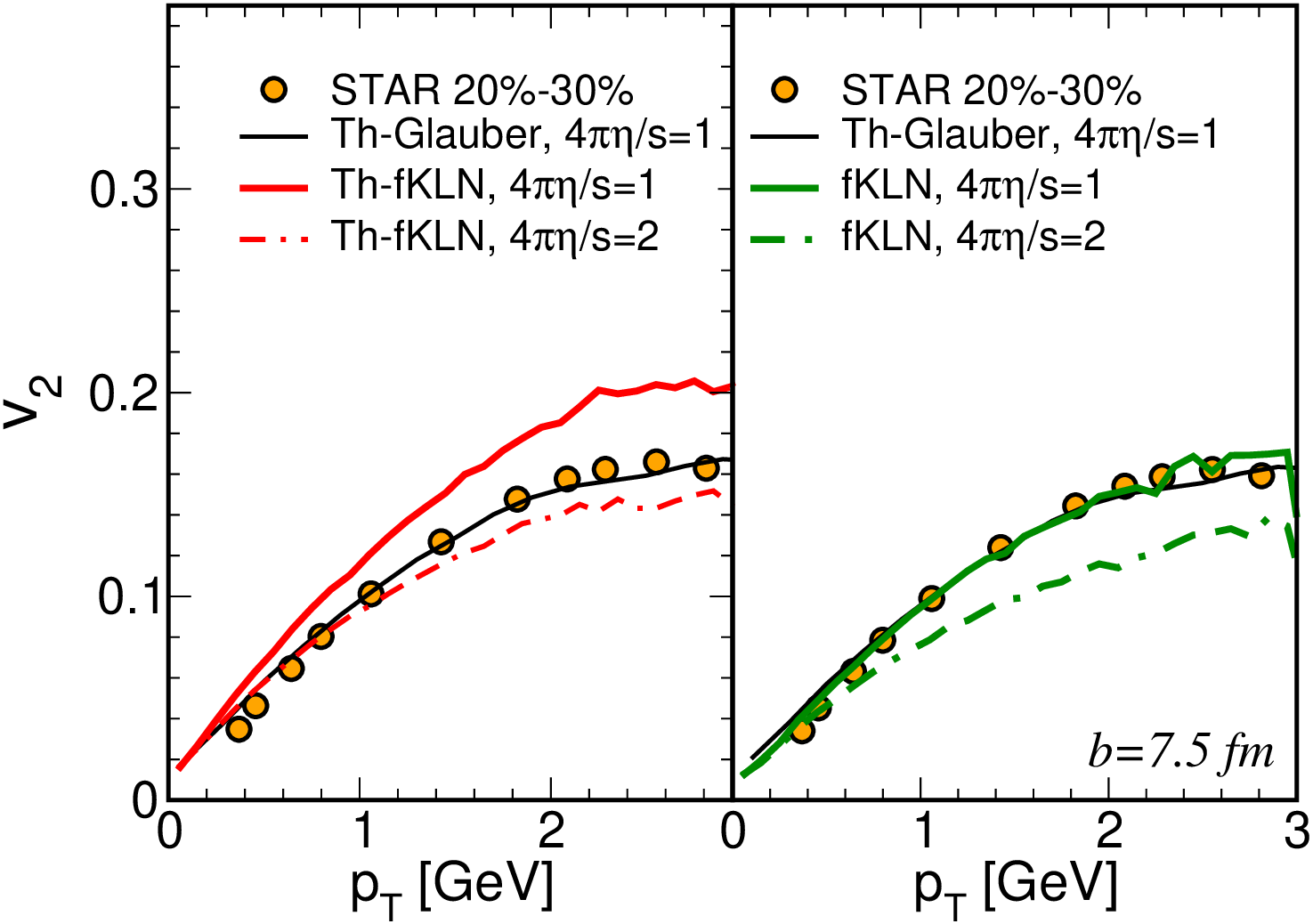}~~~\includegraphics[width=7.5cm]{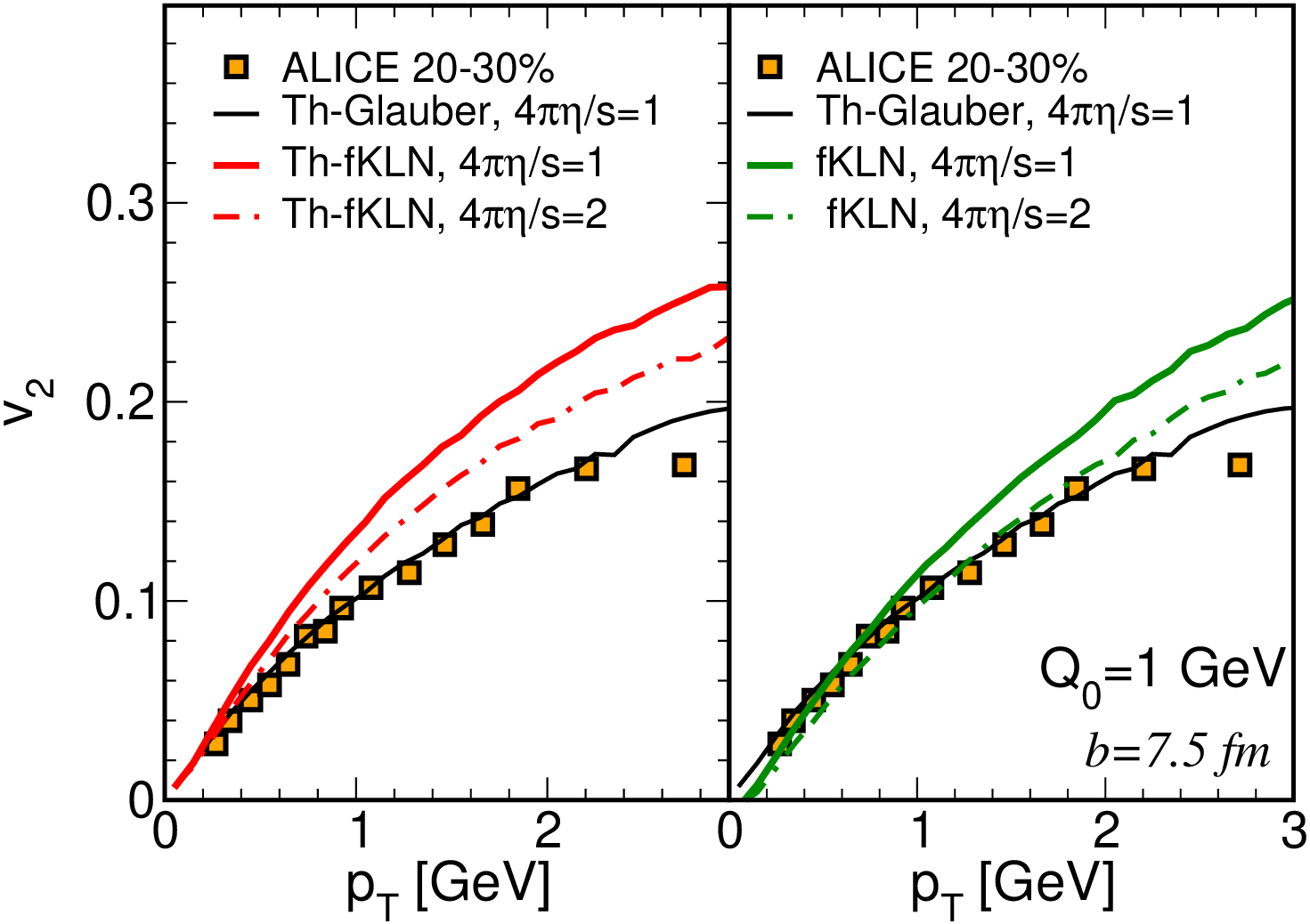}
\caption{Elliptic flow $v_2(p_T)$ at midrapidity $|y|<0.5$ for different initial conditions 
and $\eta/s$ as in the legend.
All the calculations refer to Au-Au collisions
at $\sqrt{s} = 200$ GeV. From the upper to the lower panel the impact parameter 
$b=5.3$ fm, $b=7.5$ fm and $b=9$ fm respectively.
Adapted from Ref.~\cite{Ruggieri:2013ova}.}
\label{Fig:v2pt}
\end{center}
\end{figure}

In the left panel of Fig.~\ref{Fig:v2pt}, we collect our results for the differential 
elliptic flow for the case of a Au-Au collision at RHIC energy
with $b=7.5$ fm (left panel). In the right panel we plot the 
same quantity for the case of Pb-Pb collisions at LHC energy and same impact parameter.
To guide the eye, in the figure we also plot experimental data for $v_2$ in
the relevant centrality class~\cite{Adams:2004bi}.
Since no hadronization process is yet included in our approach, the comparison 
with the charged hadrons $v_2$ has to be taken with care, even if 
it indicates that the azimuthal 
asymmetries generated by RBT are in the correct range.

Firstly we focus on the thermalized initializations, Th-Glauber and Th-fKLN.
In the case of Au-Au collisions at RHIC energy, 
the $v_2$ obtained by Th-Glauber with $4\pi\eta/s=1$
is obtained by Th-fKLN with $4\pi\eta/s \approx 2$,
because momentum distribution is the same but initial eccentricity of Th-fKLN is
larger than the one of the Th-Glauber initial condition, requiring a larger
viscosity in the former case to damp the flow.  
These results are in agreement with the ones obtained from viscous hydrodynamics 
\cite{Song:2011hk,Adare:2011tg,Luzum:2008cw}, showing the solidity and
consistency of our transport approach at fixed $\eta/s$.

Next we discuss the result for the fKLN model, represented by the green lines 
in  Fig.~\ref{Fig:v2pt},
with the proper distribution function implemented 
in both the $\bm x$ and $\bm p$ spaces. 
We find that fKLN with a $4\pi\eta/s=1$ gives a $v_2(p_T)$ quite similar
to the Th-Glauber, even if the initial eccentricity in this case is larger. 
For fKLN with $4\pi\eta/s=2$ the differential elliptic flow would be too small.
In other words the initial out-of-equilibrium fKLN distribution reduces the efficiency
in converting $\epsilon_x$ into $v_2$. 
Our interpretation is that the initial large eccentricity of the fKLN configuration
is compensated by the key feature of an almost saturated initial distribution 
in $\bm p$-space below the saturation scale $Q_s$ and probably by the softer tail at 
$p_T>Q_s$.  
We obtain similar results at the LHC energy.
In this case the Th-fKLN overestimate the Th-Glauber $v_2$ at both $4\pi\eta/s=1$ and $2$.
When the full fKLN is implemented we can see that the damping effect already discussed at RHIC
is such that with an $4\pi\eta/s=2$ one is closer to the Th-Glauber with $4\pi\eta/s=1$.

\begin{figure}[t!]
\begin{center}
\includegraphics[width=7.5cm]{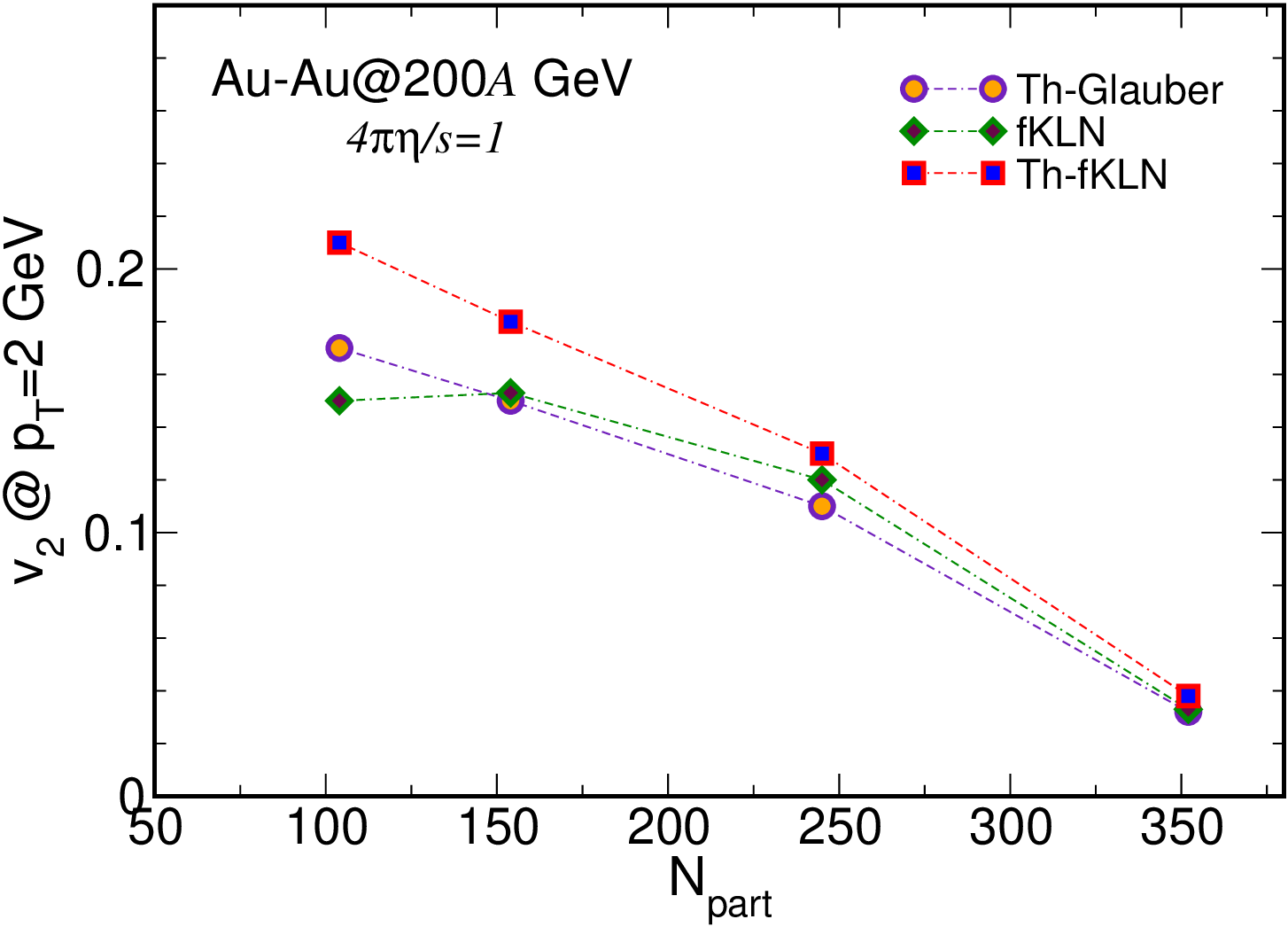}
\caption{Elliptic flow $v_2(p_T)$ at midrapidity $|y|<0.5$ and at $p_T=2$ GeV
for different initial conditions, computed at $4\pi\eta/s=1$.
All the calculations refer to Au-Au collisions
at $\sqrt{s} = 200$ GeV. Adapted from Ref.~\cite{Ruggieri:2013ova}.}
\label{Fig:v2riass}
\end{center}
\end{figure}

The effect we find on $v_2$ is more important for peripheral collisions.
In fact in Fig.~\ref{Fig:v2riass} we plot $v_2$ 
at $p_T=2$ GeV for different initializations, as a function of the number
of participants. This figure permits to visualize and summarize the dependence of $v_2(p_T)$
on the centrality class, comparing the impact of the 
initial distribution on the final $v_2$. We notice that the discrepancy between
Th-fKLN and fKLN initializations becomes less relevant for more central collisions,
implying that the effect of the initial momentum distribution is not negligible
if one considers non-central collisions.  
We also note that for central collisions at RHIC Th-Glauber and Th-fKLN for $4\pi\eta/s=1$
predict the same  $v_2$ and the effect is of KLN generating larger $v_2$ disappears.
This is seen also in viscous hydro simulation and it is a further confirmation that our approach
converge to viscous hydro  when the same thermal initial conditions are employed.

\section{Conclusions}
In this talk we have reported our results \cite{Ruggieri:2013bda,Ruggieri:2013ova} on thermalization
and building-up of the elliptic flow 
for fireballs produced in relativistic heavy ion collisions both at 
RHIC and LHC energies. We have put emphasis on the role of a nonequilibrium initial
condition on the generation of the collective flow, when the KLN initialization 
is properly implemented in coordinate and momentum space. 
Our study is based on kinetic theory at fixed $\eta/s$.
Fixing the shear viscosity over entropy density ratio permits to forget about 
microscopic processes which give that specific value of the ratio itself,
and to focus on the effect of shear viscosity on thermalization,
isotropization and building of the collective flows.

For what concerns thermalization, in both RHIC and LHC runs we have found that
thermalization times are $\tau_{therm} \approx$ 0.8-1 fm/c. 
We have then focused our attention on the elliptic flow production
when the initial distribution has a saturation scale built in it.
We have found that the amount of elliptic flow produced in heavy ion collisions
depends not only on the pressure gradients and the $\eta/s$ of the system, 
but also on the initial distribution in momentum space. 
In particular, an initial condition characterized by a momentum distribution 
with a saturation scale generates smaller $v_2$ respect to the thermal one.
Assuming the fKLN distribution as the one arising from the shattering of the CGC,
the effect of the initial nonequilibrium distribution affects the estimate of $\eta/s$ 
of about a factor of two. 
However we have also that this effect is maximal for semi-peripheral collisions,
becoming quite small for very central collisions.

In order to make more precise comparison with 
experimental data we are currently implementing also fluctuating initial conditions 
which will allow to extend the present study to all the $v_n$ harmonics relevant in HIC's.
This will allow to see if fKLN can account for the measured $v_3$ or the non-equilibrium 
damps such harmonics even more than what seen in hydrodynamics. 
In~\cite{Gale:2012rq} it is found that harmonics up to the fifth order
can be reproduced by combining CYM early-time with hydro late-evolution
evolutions; however in the calculations of~\cite{Gale:2012rq} 
a small deviation from equilibrium has to be assumed in order to 
use viscous hydrodynamical equations. Therefore it will be interesting to compute
the higher order harmonics combining the CYM initial spectrum 
with the dynamics embedded in the kinetic equations where the assumption
of sudden thermalization can be relaxed,
which will be the subject of future studies.

\ack V. G. and F. S.
acknowledge the ERC-STG funding under the QGPDyn
grant.

\end{document}